\documentclass[a4paper,11pt]{article}
\usepackage{latexsym}
\usepackage{feynmf}		%Package for feynman diagrams.
\def\ba{\begin{eqnarray}}
\def\ea{\end{eqnarray}}

\def\lb{\label}
\def\be{\begin{equation}}
\def\ee{\end{equation}}

\begin{document}

\title{General aspects of Gauss-Bonnet models without potential in dimension four}
 \author{Osvaldo P. Santill\'an \thanks{Departamento de Matem\'aticas Luis Santal\'o (IMAS), Buenos Aires, Argentina
firenzecita@hotmail.com and osantil@dm.uba.ar.}}

\date {}
\maketitle

\begin{abstract}
In the present work, the isotropic and homogenous solutions with spatial curvature $k=0$ of four dimensional Gauss-Bonnet models are characterized.
The main assumption is that the scalar field $\phi$ which is coupled to the Gauss-Bonnet term has no potential \cite{kanti1}-\cite{kanti2}. Some singular and some eternal solutions are described. The evolution 
of the universe is given in terms of a curve $\gamma=(H(\phi), \phi)$ which is the solution of a polynomial equation $P(H^2, \phi)=0$ with $\phi$ dependent coefficients.
In addition, it is shown that the initial conditions in these models put several restrictions on the evolution. For instance, an universe initially contracting will  be contracting always 
for future times and an universe that is expanding was always expanding at past times. Thus, there are no cyclic cosmological solutions for this model. These results are universal, that is,  independent on the form of the coupling $f(\phi)$ between the scalar field and the Gauss-Bonnet term. In addition, a proof that at a turning point $\dot{\phi}\to0$ a singularity necessarily emerges is presented. This is valid unless the Hubble constant $H\to 0$ at this point. This proof is based on the Raychaudhuri equation for the model. The description presented here is in part inspired in the works \cite{singularityfree}-\cite{singularityfree2}. However, the 
 mathematical methods that are implemented are complementary
 of those in these references, and they may be helpful for study more complicated situations in a future.

\end{abstract}

\section{Introduction}

One of the main interests in higher derivative gravity theories is that they can describe  inflation
by a higher order curvature in the Einstein-Hilbert action \cite{capo1}-\cite{capo2} 
without the addition of dark energy or scalar fields. An important role in this context is played by the Gauss-Bonnet invariant, since it appears in QFT renormalization in curved space times
\cite{birrell}. In addition, the Gauss-Bonnet term arises in low-energy effective actions of some string theories.
For instance, the tree-level string effective action has been calculated up to several orders in
the $\alpha'$ expansion  in \cite{3}-\cite{4}. The result is that 
there is no moduli dependence of the tree-level couplings. However, one loop corrections
to the gravitational couplings have been considered in the context of
orbifold compactifications of the heterotic superstring \cite{5}. It has been shown in that reference that
there are no moduli dependent corrections to the Einstein term while there are nontrivial
curvature contributions. They appear as the Gauss-Bonnet combination multiplied
by a function of the modulus field. 

The results described above partially motivated the study of cosmological consequences of the 
Gauss-Bonnet term. In four dimensions, this term is topological and does not
have any dynamical effect. However, when this term is non-minimally coupled
with any other field such as a scalar field $\phi$, the resulting dynamics is non trivial.
Several cosmological consequences has been exploited in recent literature, and we refer the reader
to \cite{s1}-\cite{Granda2} and references therein. But the aim of the present letter is not focused in inflationary
aspects of the theory, instead in the characterization of singular and eternal solutions of the theory.
It is important to mention that there exist preliminary works on this subject, examples are given in \cite{singularityfree}-\cite{singularityfree08}.
In particular, the results of \cite{singularityfree}-\cite{singularityfree2} suggest the existence of singular solutions as well as regular solutions. 
The singular solutions are confined to an small portion of the phase space, while the non singular fill the rest. This situation
is different than in GR, where the Gauss-Bonnet term is absent, and the powerful Hawking-Penrose theorems apply \cite{hp}.

Despite the existence of the results \cite{singularityfree}-\cite{singularityfree08}, in the present paper an independent characterization
of the possible solutions of the theory will be presented. The main assumption is that the universe is isotropic and homogeneous, that a potential is absent \cite{kanti1}-\cite{kanti2}, and that the 
spatial curvature is flat. In the authors opinion,  the mathematical characterization achieved below has not been employed yet, and it may be helpful in order to understand more complicated 
situations later on.

The present paper is organized as follows. In section 2 the equations of motion for Gauss-Bonnet vacuums in four dimensions coupled to an scalar field $\phi$ will be presented.
In particular, the full system for an isotropic and homogeneous solution with flat spatial curvature will be derived. This section do not contain any new material, all its content is standard.
 In section 3 it will be shown that the possible vacuums of the theory
are given by a curve $\gamma=(H(\phi), \phi)$ in the space $(H, \phi)$ characterized by an equation $P(H^2, \phi)=0$ which is polynomial on $H^2$. The relation is purely algebraic.
The time dependence of these fields is also derived in this section. In section 4, some numerical aspects related to these solutions are analyzed. In particular, eternal and singular solutions
are characterized. The algebraic relation found in section 3 plays an important role in this characterization. Section 5 contains a discussion of the obtained results.

\section{Gauss-Bonnet equation}

The model that will be considered here is a generalization of the Gauss-Bonnet one. Recall that a pure Gauss-Bonnet gravity model is described in
D dimensions by the following action 
\begin{equation}
S=\int d^{D}x \sqrt{-g}G.
\label{accionGB}
\end{equation}
The Gauss-Bonnet invariant $G$ is given by
\begin{equation}
G\equiv R^2-4R_{\alpha \beta}R^{\alpha \beta}+R_{\alpha \beta\gamma\delta}R^{\alpha \beta\gamma\delta}.
\label{gaussbonnet1}
\end{equation}
The signature to be employed in the following is $(-, +, ..., +)$.
The equations of motions $\delta S=0$ that arise by considering variations $\delta g_{\mu\nu}$ are the following
$$
-\frac{1}{2}R_{\rho\sigma}R^{\rho\sigma}g_{\alpha\beta}-\nabla_{\alpha}\nabla_{\beta}R-2R_{\rho\beta\alpha\sigma}R^{\sigma\rho}+\frac{1}{2}g_{\alpha\beta}\Box R
+ \Box R_{\alpha\beta}+\frac{1}{2}R^{2}g_{\alpha\beta}-2RR_{\alpha\beta}
$$
\be
-2\nabla_{\beta}\nabla_{\alpha}R+2g_{\alpha\beta}\Box R -\frac{1}{2}R+R_{\alpha\beta}=0.
\label{gb}
\end{equation}
The resulting model is non trivial for $D\neq 4$. In four dimensions instead, the term $\sqrt{-g}G$ can be expressed as a total derivative
\begin{equation}
\sqrt{-g}G=\partial_{\alpha}K^\alpha,
\qquad 
K^\alpha=\sqrt{-g}\epsilon^{\alpha\beta\gamma\delta}\epsilon_{\rho\sigma}^{\mu\nu}\Gamma^{\rho}_{\mu\beta}\bigg[\frac{R^{\sigma}_{\nu\gamma\delta}}{2}+\frac{\Gamma^{\sigma}_{\lambda\gamma}\Gamma^{\lambda}_{\nu\sigma}}{3}\bigg].
\label{lalala}
\end{equation}
Therefore in $D=4$ and with a manifold without boundary, this model is irrelevant.
However, the following modified action
\begin{equation}\label{g2}
S=\int d^4x \sqrt{-g}\left\{ \frac{1}{2\kappa^2}R
    - \frac{1}{2}\partial_\mu \phi \partial^\mu \phi+V(\phi)+ f(\phi) G\right\},
\end{equation}
is physically non trivial, as the Gauss-Bonnet is coupled to the real scalar field $\phi$ by the coupling $f(\phi)$. Due to this coupling, the lagrangian it is not a total derivative anymore and contributes to the equations of motion. 

In the following, a Gauss-Bonnet model with $V(\phi)=0$ will be considered, as in references \cite{kanti1}-\cite{kanti2}. The equation for the scalar field $\phi$ under this circumstance is given by
\begin{equation}
\label{g3}
\nabla^2 \phi+ f'(\phi) G=0.
\end{equation}
The equations of motion for the metric $g_{\mu\nu}$ are more involved. The variation of the action throws the following result
$$
0=\frac{1}{\kappa^2}\left(- R^{\mu\nu} + \frac{1}{2} g^{\mu\nu} R\right)+\frac{1}{2}\partial^\mu \phi \partial^\nu \phi
 - \frac{1}{4}g^{\mu\nu} \partial_\rho \phi \partial^\rho \phi+ \frac{1}{2}g^{\mu\nu}f(\phi) G 
-2 f(\phi) R R^{\mu\nu} 
$$
$$
+ 2 \nabla^\mu \nabla^\nu \left(f(\phi)R\right)- 2 g^{\mu\nu}\nabla^2\left(f(\phi)R\right) 
+ 8f(\phi)R^\mu_{\ \rho} R^{\nu\rho}- 4 \nabla_\rho \nabla^\mu \left(f(\phi)R^{\nu\rho}\right)
 - 4 \nabla_\rho \nabla^\nu \left(f(\phi)R^{\mu\rho}\right)
 $$
 $$
 + 4 \nabla^2 \left( f(\phi) R^{\mu\nu}  \right)+ 4g^{\mu\nu} \nabla_{\rho} \nabla_\sigma \left(f(\phi) R^{\rho\sigma} \right)
- 2 f(\phi) R^{\mu\rho\sigma\tau}R^\nu_{\ \rho\sigma\tau}+ 4 \nabla_\rho \nabla_\sigma \left(f(\phi) R^{\mu\rho\sigma\nu}\right).
$$
However, by taking into account the following identities
$$
\nabla^\rho R_{\rho\tau\mu\nu}= \nabla_\mu R_{\nu\tau} - \nabla_\nu
R_{\mu\tau},
$$
$$
\nabla^\rho R_{\rho\mu} = \frac{1}{2} \nabla_\mu R,
$$
$$
\nabla_\rho \nabla_\sigma R^{\mu\rho\nu\sigma} =
\nabla^2 R^{\mu\nu} - {1 \over 2}\nabla^\mu \nabla^\nu R
+ R^{\mu\rho\nu\sigma} R_{\rho\sigma}- R^\mu_{\ \rho} R^{\nu\rho},
$$
$$      
\nabla_\rho \nabla^\mu R^{\rho\nu}
+ \nabla_\rho \nabla^\nu R^{\rho\mu}
= {1 \over 2} \left(\nabla^\mu \nabla^\nu R
+ \nabla^\nu \nabla^\mu R\right)
 - 2 R^{\mu\rho\nu\sigma} R_{\rho\sigma}
+ 2 R^\mu_{\ \rho} R^{\nu\rho},
$$
$$
\nabla_\rho \nabla_\sigma R^{\rho\sigma} = \frac{1}{2} \Box R,
$$
which are consequences of the Bianchi identities, the last expression can be written as \cite{Nojiri}
$$
0=\frac{1}{\kappa^2}\left(- R^{\mu\nu} + \frac{1}{2} g^{\mu\nu} R\right)
      +  \left(\frac{1}{2}\partial^\mu \phi \partial^\nu \phi
      - \frac{1}{4}g^{\mu\nu} \partial_\rho \phi \partial^\rho \phi \right)
   + \frac{1}{2}g^{\mu\nu}f(\phi) G
$$   $$  
   -2 f(\phi) R R^{\mu\nu} + 4f(\phi)R^\mu_{\ \rho} R^{\nu\rho}
   -2 f(\phi) R^{\mu\rho\sigma\tau}R^\nu_{\ \rho\sigma\tau}
+4 f(\phi) R^{\mu\rho\sigma\nu}R_{\rho\sigma} 
$$
$$ 
+ 2 \left( \nabla^\mu \nabla^\nu f(\phi)\right)R
      - 2 g^{\mu\nu} \left( \nabla^2f(\phi)\right)R
   - 4 \left( \nabla_\rho \nabla^\mu f(\phi)\right)R^{\nu\rho}
      - 4 \left( \nabla_\rho \nabla^\nu f(\phi)\right)R^{\mu\rho} 
$$
\begin{equation}
\label{gb4b}
+ 4 \left( \nabla^2 f(\phi) \right)R^{\mu\nu}
+ 4g^{\mu\nu} \left( \nabla_{\rho} \nabla_\sigma f(\phi) \right) R^{\rho\sigma}
- 4 \left(\nabla_\rho \nabla_\sigma f(\phi) \right) R^{\mu\rho\nu\sigma}.
\ee
The equations (\ref{g3}) and (\ref{gb4b}) are the full system of equations describing the theory.

The following discussion is focused on the isotropic and homogeneous vacuums of the model 
with zero spatial curvature. The corresponding  distance element for these vacuums is given by
$$
g_4=-dt^2+a^2(t)\sum_{i=1}^3 dx_i^2.
$$
The formulas for the  Levi-Civita connection and the curvature of this background are well known, they are explicitly
$$
\Gamma^t_{ij}=a^2 H \delta_{ij}\ ,\qquad
\Gamma^i_{jt}=\Gamma^i_{tj}=H\delta^i_{\ j}\ ,
\qquad R_{itjt}=-\left(\dot H + H^2\right)\delta_{ij},
$$
$$
R_{ijkl}=a^4 H^2\left(\delta_{ik} \delta_{lj} - \delta_{il} \delta_{kj}\right),\qquad
R_{tt}=-3\left(\dot H + H^2\right)\ ,
$$
\be\lb{wk}
R_{ij}= a^2 \left(\dot H
+ 3H^2\right)\delta_{ij},\qquad R= 6\dot H + 12 H^2.
\ee
The other components are all zero. By use of these formulas,  equation (\ref{g3}) becomes
\be\lb{uco}
\ddot{\phi}+3H\dot{\phi}-24H^2 f'(\phi)(H^2+\dot{H})=0.
\ee
The two other independent equations that follows from (\ref{gb4b}) are 
\be\lb{ico}
H^2=\frac{\kappa^2}{3}\rho_{eff},\qquad 2\dot{H}+3H^2=-\kappa^2 p_{eff},
\ee
where the energy density and the pressure of the scalar field $\phi$ are given by
$$
\rho_{eff}=\frac{\dot{\phi}^2}{2}-24 H^3 \dot{f},\qquad p_{eff}=\frac{\dot{\phi}^2}{2}+8H^2 f''(\phi)\dot{\phi}^2+8H^2 f'(\phi)\ddot{\phi}+16 H\dot{H} f'(\phi)\dot{\phi}+16 H^3 f'(\phi)\dot{\phi},
$$
respectively. The equations (\ref{uco}) and (\ref{ico}) characterize the isotropic and homogeneous vacuums of the theory, with zero spatial curvature.

\section{Formal aspects}
The next task is to characterize the possible solutions of the system (\ref{uco})-(\ref{ico}). By a redefinition of the scalar field $\phi$ and the function $f(\phi)$ the Newton constant
$\kappa$ may be set to one. In addition the redefinition $8f(\phi)\to f(\phi)$ is convenient, in order to simplify some numerical factors. 
After these redefinitions, the equations of motion (\ref{uco}) and (\ref{ico}) are given by
\be\lb{1}
\frac{\dot{\phi}^2}{2}=3H^2(1+\dot{f}(\phi)H),
\ee
\be\lb{2}
\frac{\dot{\phi}^2}{2}=-2(H^2+\dot{H})(1+\dot{f}(\phi)H)-H^2(1+\ddot{f}(\phi)),
\ee
\be\lb{3}
\ddot{\phi}=-3H\dot{\phi}+3H^2 f'(\phi)(H^2+\dot{H}).
\ee
This is a highly non linear system for $\phi$ and $H$. However, several conclusions may be drawn from (\ref{1})-(\ref{3}) without reference to a particular choice of the coupling $f(\phi)$. First,
by taking into account that $\dot{f}(\phi)=f'(\phi)\dot{\phi}$,  the equation (\ref{1}) becomes a quadratic algebraic relation for $\dot{\phi}$.
Its solution is 
\be\lb{c1}
\dot{\phi}=\frac{H}{2}\bigg[6H^2f'(\phi)\pm \sqrt{36H^4f'(\phi)^2+24}\bigg].
\ee
This equation, when inserted into (\ref{3}), gives the following expression for $\ddot{\phi}$
\be\lb{c3}
\ddot{\phi}=-\frac{3}{2}H^2\bigg[6H^2f'(\phi)\pm \sqrt{36H^4f'(\phi)^2+24}\bigg]+3 H^2 f'(\phi)(H^2+\dot{H}).
\ee
On the other hand, a different expression for $\ddot{\phi}$ can be obtained by directly taking the time derivative of (\ref{c1}), the result is
$$
\ddot{\phi}=\frac{\dot{H}}{2}\bigg[18 H^2 f'(\phi)\pm \frac{216 H^5 f'(\phi)^2+48H}{\sqrt{36H^6f'(\phi)^2+24H^2}}\bigg]
$$
\be\lb{bravado}
+\frac{H}{2}\bigg[6H^2f'(\phi)\pm \sqrt{36H^4f'(\phi)^2+24}\bigg]\bigg[6 H^3 f''(\phi)\pm \frac{72 H^6 f'(\phi)f''(\phi)}{\sqrt{36H^6f'(\phi)^2+24H^2}}\bigg].
\ee
 Both expressions (\ref{bravado}) and (\ref{c3}) should be equal, and this equality implies that
$$
\bigg[6H^2f'(\phi)\pm \frac{216 H^5 f'(\phi)^2+48H}{2\sqrt{36H^6f'(\phi)^2+24H^2}}\bigg]\dot{H}=-\frac{3}{2}H^2\bigg[6H^2f'(\phi)\pm \sqrt{36H^4f'(\phi)^2+24}\bigg]
$$
\be\lb{c2}
+3H^4 f'(\phi)-\bigg[3H^2f'(\phi)\pm \sqrt{9H^4f'(\phi)^2+6}\bigg]\bigg[6H^4f''(\phi)\pm \frac{72H^7f'(\phi)f''(\phi)}{\sqrt{36H^4f'(\phi)^2+24}}\bigg].
\ee
This is an equation which expresses the time derivative  $\dot{H}$ of the Hubble parameter as a function $\dot{H}=\dot{H}(\phi, H)$. 
However, the formula (\ref{2}) was not used to get this relation. When this formula is taken into account, a non equivalent expression for $\dot{H}=H(\phi, H)$ is obtained as follows. By use of the elementary formula
$\ddot{f}=f''(\phi)\dot{\phi}^2+f'(\phi)\ddot{\phi}$ together with (\ref{1}) and (\ref{c1}), the formula (\ref{2}) can be expressed as follows
\be\lb{cc2}
-H^2f'(\phi)\ddot{\phi}=H^2+\bigg[1+\frac{H^2}{2}f'(\phi)\bigg[6H^2f'(\phi)\pm \sqrt{36H^4f'(\phi)^2+24}\bigg]\bigg](5H^2+2\dot{H}+6H^4 f''(\phi)).
\ee
By replacing $\ddot{\phi}$ by (\ref{c3}) it follows after some simple rearrangement of terms that  
$$
\bigg\{2+H^2f'(\phi)\bigg[3H^2f'(\phi)\pm \sqrt{36H^4f'(\phi)^2+24}\bigg]\bigg\}\dot{H}=-H^2
$$
$$
-\frac{3H^4f'(\phi)}{2}\bigg[6H^2f'(\phi)\pm \sqrt{36H^4f'(\phi)^2+24}\bigg]
+3H^6f'(\phi)^2-H^2
$$
\be\lb{cc3}
-\frac{1}{2}\bigg[2+H^2f'(\phi)\bigg[6H^2f'(\phi)\pm \sqrt{36H^4f'(\phi)^2+24}\bigg]\bigg](5H^2+6H^4 f''(\phi)),
\ee
which is an expression for $\dot{H}=\dot{H}(\phi, H)$ non equivalent to (\ref{c2}).

Now, the crucial point is that the non equivalent expressions
(\ref{c2}) and (\ref{cc3}) are of the form
$$
A \dot{H}=B,\qquad C\dot{H}=D,
$$
respectively, where $A,.., D$ are functions of $H$ and $\phi$ but not functions of $\dot{H}$. The compatibility conditions
for both equations to be true is $AD=BC$. By reading the coefficients $A, ..., D$ directly from  (\ref{c2}) and (\ref{cc3})  it follows
that the relation $AD=BC$ translates into an equation of the form
$$
(a H^{10}f'(\phi)^4f''(\phi)+b H^8f'(\phi)^4+c H^{6}f'(\phi)^2f''(\phi)+d H^4f'(\phi)^4+e H^{2}f'(\phi)+f) H^\alpha
$$
\be\lb{compol}
=\bigg[g H^8f'(\phi)^4+h H^{6}f'(\phi)^2f''(\phi)+m H^4f'(\phi)^4+n H^{2}f'(\phi)+q\bigg] H^\alpha \sqrt{36H^4f'(\phi)^2+24}.
\ee
Here $a,..,q$ are some real numbers and $\alpha>0$ is an integer, whose explicit value is not of importance in the following discussion.
The factor $H^\alpha$ is important, since it shows that $H=0$ may be a solution of this equation. In fact, when $H\to 0$ both equations (\ref{c2}) and (\ref{cc3})
show that, if $f(\phi)$ and its derivatives are non divergent, then $\dot{H}\to0$. Thus, the flat space is included in the model.

The solutions of (\ref{compol}) give a relation between $H$ and $\phi$, which is of course non univoque. By taking the
square of this expression it follows that these branches $H(\phi)$ satisfy 
$$
H^{2\alpha}\bigg[a H^{10}f'(\phi)^4f''(\phi)+b H^8f'(\phi)^4+c H^{6}f'(\phi)^2f''(\phi)+d H^4f'(\phi)^4+e H^{2}f'(\phi)+f\bigg]^2
$$
\be\lb{compo}
-H^{2\alpha}\bigg[g H^8f'(\phi)^4+h H^{6}f'(\phi)^2f''(\phi)+m H^4f'(\phi)^4+n H^{2}f'(\phi)+q\bigg]^2 (36H^4f'(\phi)^2+24)=0.
\ee
This is  an algebraic relation of the form $P(H^2, f'(\phi), f''(\phi))=0$, which is polynomial of degree ten for $H^2$ (or an expression
for $H$ of order twenty), up to the factor $H^{2\alpha}$. The equation $P(H^2,  f'(\phi), f''(\phi))=0$ describe a curve $\gamma=(H(\phi), \phi)$ in the space
$(H, \phi)$ of possible evolutions of the vacuum, without reference to the proper time $t$. This should be supplemented with the  time dependence of some 
of the fields $\phi$ or $H^2$. This dependence follows directly from (\ref{c1}), the result is 
\be\lb{compo2}
t-t_0=\int_{\phi_0}^{\phi} \frac{d\phi}{H\bigg[6H^2f'(\phi)\pm \sqrt{36H^4f'(\phi)^2+24}\bigg]},
\ee
and is obtained by simple integration.

Both expressions (\ref{compo})-(\ref{compo2}) represent the full solution of the Einstein system in the isotropic and homogeneous case. The first gives  implicitly $H=H(\phi)$ and this, combined with the second gives that $\phi=\phi(t)$ and consequently $H=H(t)$. In the general case, an explicit expression for the solution is hopeless by the Galois theorem. Nevertheless, as it will be explained below, this algebraic relation will be useful for  partially describing the solutions of the theory.

\section{Numerical aspects}
\subsection{Bounds for the evolution}
In addition to the formal characterization made above, some numerical consequences of the model may found as follows. First of all, from equation (\ref{1}) it is immediately deduced that
\be\lb{polo}
\frac{\dot{\phi}^2}{6H^2}=1+\dot{f}H\geq0.
\ee
This implies that the inequality
$$
H f'(\phi)\dot{\phi}\geq-1,
$$
is satisfied during the whole evolution of the universe. On the other hand, the equation (\ref{2}) can be expressed by use of (\ref{1}) as follows
$$
\frac{5\dot{\phi}^2}{6}+H^2=-2\frac{dH}{dt}-\frac{d(\dot{f}H^2)}{dt}\geq 0.
$$
Therefore, as the proper time $t$ grows, the following bound is obtained
\be\lb{ready}
H(2+\dot{f}H)\leq C_0,\qquad t\geq0,
\ee
with $C_0$ being the  value of the quantity $H(2+\dot{f}H)$ at $t=0$. 

The inequality (\ref{ready}) derived above has important consequences. 
Assume for a moment that the initial condition is $C_0<0$. The inequality (\ref{polo}) shows that the factor $2+\dot{f}H\geq 1$, therefore
this case corresponds to $H< 0$. Thus the universe is contracting at $t=0$. By taking into account (\ref{ready}) and that $2+\dot{f}H\geq 1$ it follows that
\be\lb{ready2}
H\leq \frac{C_0}{2+\dot{f}H}\leq 0,\qquad t\geq0.
\ee
Thus the universe is always contracting in the future if it is contracting at $t=0$. This is an universal conclusion, no matter the form of the coupling 
$f(\phi)$.
Furthermore, for the past, the equation (\ref{ready}) is converted into
\be\lb{readyatras}
H\geq \frac{C_0}{2+\dot{f}H}\geq C_0,\qquad t\leq0,
\ee
Again, in the last step the inequality $2+\dot{f}H\geq 1$ was taken into account.

Suppose now that  the initial condition is $C_0>0$. Then
\be\lb{ready2}
H\leq \frac{C_0}{2+\dot{f}H}\leq C_0,\qquad t\geq0.
\ee
For the past, the equation (\ref{ready}) is converted into
\be\lb{readyatras}
H\geq \frac{C_0}{2+\dot{f}H}\geq 0,\qquad t\leq0,
\ee
Thus, if the universe is expanding at $t=0$, it was expanding always in the past. Again, this is an universal conclusion, without reference to the particular
form of $f(\phi)$. These results show in particular that there are no cyclic cosmologies for these models.

Consider now the behavior of $\phi$ and $\dot{\phi}$. If the negative branch is chosen in (\ref{c1}) then
the time derivative is given by
\be\lb{c1d}
2\dot{\phi}=H\bigg[6H^2f'(\phi)-\sqrt{36H^4f'(\phi)^2+24}\bigg].
\ee
It is convenient to express this formula as
\be\lb{c5d}
2\dot{\phi}=6H^3f'(\phi)\bigg[1-\sqrt{1+\frac{24}{36H^4f'(\phi)^2}}\bigg].
\ee
In these terms it follows from (\ref{c5d}) that if $f'(\phi)\to\pm\infty$ then $\dot{\phi}\to 0$. Also, if $f'(\phi)\neq 0$ then $\dot{\phi}\to 0$ when $H\to\pm\infty$.
Instead, if $f'(\phi)=0$ then $\dot{\phi}\to\pm\infty$ when $H\to \pm \infty$, as follows from (\ref{c1d}). For this reason, a coupling $f'(\phi)$ 
which never reaches a zero will  be chosen. An example of this may be a coupling $f'(\phi)>0$ with a minimum $f'_m>0$. If this condition is satisfied, $\dot{\phi}$ is never divergent. In addition
$\dot{\phi}\to 0$ when $H\to 0$. 

Now, if $\dot{\phi}$ is interpreted as a function of the two variables $(f'(\phi), H)$ given by (\ref{c1d}), then the properties shown above show that it vanishes
at the point $(f'_m,0)$ and at any point of the infinite. The fact that $\dot{\phi}$ is a continuous function shows that it must have a minimum and a maximum somewhere. In fact, the restriction
of this function to any straight line in the space $(f'(\phi), H)$ connecting the point $(f'_m,0)$ with some point of at the infinite interpolates between the two zeros continuously. The Bolzano theorem implies  the presence of a minimum or a maximum in any of these directions. These directions are parameterized
an "angular" coordinate $\vartheta$ and since the function $\dot{\phi}$ is well behaved,  these extrema varies continuously with the angle. As the angular coordinate $0\leq\vartheta<2\pi$ is compact,  it follows that there should exist a global minimum $\dot{\phi}_1$ and a global maximum $\dot{\phi}_2$. Therefore
$$
\dot{\phi}_2\leq\dot{\phi}\leq\dot{\phi}_1,
$$
and, by simple integration of the last expression, it follows that
$$
 \phi_0+\dot{\phi}_1t\leq\phi\leq\phi_0+\dot{\phi}_2t,\qquad t\geq 0,
$$
\be\lb{bondo}
 \phi_0+\dot{\phi}_2 t\leq\phi\leq\phi_0+\dot{\phi}_1t,\qquad t\leq 0.
\ee
This means that the values of $\phi$ are bounded by two linear time functions, and therefore are bounded for any finite time $t$.

\subsection{Characterization of the singular and regular solutions}

In order to characterize the regular and singular solutions of the theory, it should be remarked that the curvature invariants that can be constructed by (\ref{wk})
are all dependent on $H$ and its derivative $\dot{H}$. For instance, the following invariants
$$
R= 6\dot H + 12 H^2,\qquad R_{ij}R^{ij}=12 \left(\dot H
+ 3H^2\right)^2,
$$
are solely dependent on these two quantities. The following discussion is focused in the possible singularities of $H$ and $\dot{H}$.

The bounds described in the previous subsection are useful in order to understand these singularities. First, note that the equation (\ref{compo}) can be considered as an algebraic one with coefficients
determined the first and second derivatives $f'(\phi)$ and $f''(\phi)$. If these functions exist and are continuous for any finite value of $\phi$ and are never zero, then it is obtained from (\ref{compo}) 
and (\ref{bondo})
that these coefficients  $f'(\phi(t))$ and $f''(\phi(t))$ are well behaved for every finite value of the proper time $t$. Of course, this assumption implies that these functions have no vertical asymptotes.
The assumption that the coefficients are never zero is for simplicity, otherwise the behavior of the roots of a polynomial may be singular when some of the coefficients vanish.\footnote{To see an example, consider a quadratic whose principal coefficient goes to zero. It is easy to see that one of the roots is divergent in this limit.}
Under these conditions, by taking into account that the coefficients are finite and are symmetric functions of the roots, it follows that the roots $H(\phi)$ of (\ref{compo}) are finite for every finite time $t$. 

The next step is to understand the possible singularities of $\dot{H}$. It is convenient to consider a simple example first. Suppose for a moment that there is a (fictitious) theory whose
vacuum is described by a circle $H^2+\phi^2=1$. Then it is clear that, at the point $(H, \phi)=(1, 0)$ the value of $H'(\phi)$ is divergent, since the tangent
to the curve is vertical to the $H$ axis. If the dynamic is such that this point $(H, \phi)=(1, 0)$ is attained at finite time, then $\dot{H}=H'(\phi)\dot{\phi}$ appears to be divergent due to the fact that
 $H'(\phi)\to\infty$ at this point. However, it is also clear that $\dot{\phi}$ goes to zero at $(H, \phi)=(1, 0)$, since it is a turning point for $\phi$. Thus a $0.\infty$ type of indetermination
 appears. But one may think a plenty of situations  such that the fields are evolving around the circle with finite velocity, for instance, $H=\sin(t)$ and $\phi=\cos(t)$. In those cases, the $0.\infty$ indetermination gives a finite answer and $\dot{H}$ is finite at the turning point.

For the present model, the situation is more complicated  than in the example just described, as the curve $\gamma=(H^2(\phi), \phi)$ is described by  (\ref{compo}) which is much harder than a circle equation.
Still, some conclusions may be obtained. The equation (\ref{compo}) for $H^2$ is of  generic form
$$
P_n(H^2)=\sum_{n=0}^{10} a_n H^{2n}=0,
$$
with $\phi$ dependent coefficients $a_n$, up to a factor $H^{2\alpha}$. This last factor describes a flat space solution. Now,  the coefficients of this equation 
are well behaved functions of $\phi$ and therefore an infinitesimal change  $d\phi$ induces an smooth change in the coefficients $da_i$. This induces a change $dH^2$ in any of the roots
of the polynomial, which is related to the variations $da_i$ by the following formula
\be\lb{deri}
P'_n(H^2) dH^2=-\sum_{n=0}^9 H^{2n} da_n.
\ee
If this formula is not satisfied, then $H^2+dH^2$ would cease to be a root.
From the last relation, it is deduced that
\be\lb{par}
\frac{\partial H^2}{\partial a_i}=-\frac{H^{2n}}{P'_n(H^2)}.
\ee
Thus the derivative (\ref{par}) is well behaved when $P_n(H^2)=0$ but $P'_n(H^2)\neq 0$. For a polynomial function, this condition is the statement that the roots $H^2$ are simple roots. In other words, if the evolution of $\phi$ is such that two roots $H^2_1$ and $H^2_2$ merge at a finite time $t_0$, then at this point the derivative (\ref{par}) will be divergent.  Now, the time derivative of $H^2$ is
\be\lb{wu}
\frac{dH^2}{dt}=\frac{\partial H^2}{\partial a_i}\frac{da_i}{d\phi}\dot{\phi}.
\ee
As was argued above, the Hubble constant $H$ has finite values at any finite time $t$. Thus the possible divergent behavior of the derivative
would not be due to a vertical asymptote. Instead, it will be due to the presence of a turning point in the curve $\gamma=(H^2(\phi), \phi)$, which is characterized by $\dot{\phi}\to0$.

It is tempting to conclude, in view of the circle example given above, that the  $0.\infty$ indetermination that appears in (\ref{wu}) at a turning point is generically solved to give a finite answer.  Therefore the resulting value of $\dot{H}$ and consequently, the value of $R$, will be finite at such point. However, this is absolutely not the case, and a rigorous proof can be given as follows. Consider the Raychaudhuri identity \cite{hp}
\be\lb{roy}
\frac{d\theta}{dt}=-R_{tt}-\frac{\theta^2}{3},
\ee
which is an useful tool for studying singularities. Here $\theta$ is the expansion parameter of the universe at a given time $t$, which in the isotropic and homogenous case reduces $\theta=3H$.  For the space times in consideration one has that
$$
R_{tt}=-3\left(\dot H + H^2\right).
$$
The insertion of this expression into (\ref{roy}) gives a trivial identity. 
However, a non trivial relation may be found by use of (\ref{ico}). By taking (\ref{ico}) into account, together with the redefinition $\kappa^2=1$ and $8f(\phi)\to f(\phi)$
the last formula becomes
$$
R_{tt}=\frac{1}{2}(\rho_{eff}+3 p_{eff})=\dot{\phi}^2+\frac{3}{2}H^2 f''(\phi)\dot{\phi}^2+\frac{3}{2}H^2 f'(\phi)\ddot{\phi}+3 H\dot{H} f'(\phi)\dot{\phi}+\frac{3}{2} H^3 f'(\phi)\dot{\phi}.
$$
This, together with identification $\theta=3H$, converts the Raychaudhuri equation (\ref{roy}) into 
\be\lb{roy2}
3(1+H\dot{f})\frac{dH}{dt}=-\frac{\dot{\phi}^2}{2}-\frac{3}{2}H^2 (\ddot{f}+H\dot{f})-3H^2.
\ee
By further use of (\ref{polo}) the last equation can be expressed as 
\be\lb{roy3}
\dot{\phi}^2\frac{dH}{dt}=-2H^2\dot{\phi}^2-3H^4(\ddot{f}+H\dot{f})-6H^4.
\ee
This equation is completely equivalent to (\ref{2}). But some indirect conclusions may be drawn from this  precise expression. First, if $H\neq 0$ then, for a turning point characterized by $\dot{\phi} \to 0$ it follows $\dot{H}\to\pm\infty$. Thus, generically for the present model, a singular value for $\dot{H}$ and for the curvature $R$ are present at a turning point. This is an important conclusion. 

In brief, the analysis given above shows  that the curvature $R$ of the universe will be regular if there are no turning points for $\phi$. If this is combined with (\ref{wu}) it follows that there is no turning point when the derivative $\partial_{a_i}H^2$ is finite. This implies that the curvature $R$ will be regular if the roots of the polynomial equation (\ref{compo}) do not merge at any finite proper time $t$.  Therefore, the task to construct universes with regular curvature $R$ is reduced to find a function $f(\phi)$ is $C^3$, such that $f'(\phi)\neq 0$ and is such that the roots of
(\ref{compo}) never merge, no matter which values $\phi$ takes during the evolution. There exist plenty of coupling functions of this type.  Consider a coupling such that  $f'(\phi)$ has a minimum $f'_1>0$ and a maximum $f'_2$, and such that these extrema are reached asymptotically at $\phi\to\pm\infty$. Furthermore assume that the second derivative $f''(\phi)$ is never zero except at $\phi\to\pm\infty$.  Then both $f'(\phi)$ and $f''(\phi)$ take bounded values, and so the coefficients of the polynomial (\ref{compo}). By adjusting the range of values of $f'(\phi)$ and $f''(\phi)$ conveniently, it should be possible to construct a model for which the roots never merge due to the limited values of the polynomial coefficients.\footnote{Note that the condition that the roots never merge is too restrictive. In fact, the condition that is really needed is that \emph{some} of the roots do not merge.}

A further possibility is that the evolution of $\phi$, which is already constrained by (\ref{bondo}), is in fact  constrained further to a compact set $\phi_1\leq\phi(t)\leq \phi_2$ for which the image of the polynomial coefficients of (\ref{compo}) determined by $f'(\phi(t))$ and $f''(\phi(t))$ are such that the roots never merge. However, without knowing the precise form of the coupling and the evolution of $\phi(t)$ we can not make further comments about this possibility.

\section{Discussion}

In the present note, a mathematical characterization of isotropic and homogeneous vacuum of scalar Gauss-Bonnet inflationary models without potential
was presented. For spatially flat universe, the solution was described in terms of a polynomial equation (\ref{compo}) for the Hubble constant $H$ with $\phi$ dependent coefficients.
In addition, it was shown that the Hubble constant $H$ is bounded by the initial condition for any future time, independently of the form of the coupling $f(\phi)$. In particular, an initially contracting universe
is contracting forever in the future, independently on the form of the coupling $f'(\phi)$. Also, an expanding universe was always expanding in the past. Thus, no cyclic cosmologies are allowed in this model.
Another result is that, at a turning point $\dot{\phi}\to 0$ the universe is necessarily singular, unless it is flat. These results are universal, that is, independent on the form of the coupling $f(\phi)$, and can be interpreted as  no-go theorem.

Other results were derived as well. However, it is important to emphasize that these further results are not a no-go theorem. For instance, it has been shown that when the derivative of the coupling $f'(\phi)$ between the inflaton and the Gauss Bonnet term is bounded from above and below, there exist non trivial cosmological solutions
whose curvature is regular for any past and future times.  However, if these extrema do not exist, it does not mean that the solution will be singular. In fact, in reference \cite{konto} some singularity free solutions with $f(\phi)=a \phi^2$ were described. Clearly, $f'(\phi)$ does not have a minimum value, and the resulting solution is still regular.

Finally, we would like to remark that, even though the absence of a potential for the inflaton may be non realistic from the cosmological point of view \cite{nogo}, the mathematical methods implemented here may be useful to characterize solutions with the potential term $V(\phi)$ or spatial curvature $k=\pm 1$ turned on. We will come to this point in a near future.

\section*{Acknowledgments}
The author is supported by the CONICET.

\end{document}